\documentclass[prd,twocolumn,showpacs,floatfix,nofootinbib,amssymb,floatfix]{revtex4}
\usepackage{graphicx,color,dcolumn,booktabs,bm,multirow,cleveref}
\usepackage{longtable,lscape}
\usepackage{txfonts}
\usepackage{overpic}
\usepackage{amssymb}
\usepackage{indentfirst}
\usepackage{feynmf}   
\usepackage{slashed}  
\usepackage{cases}
\usepackage{color}
\usepackage{multirow}
\usepackage{epstopdf}
\usepackage{graphicx,color,dcolumn,booktabs,bm}
\usepackage[colorlinks,
            citecolor=blue,
            anchorcolor=red,
            menucolor=red,
            linkcolor=red,
            filecolor=red,
            runcolor=red,
            urlcolor=blue,
            frenchlinks=red]{hyperref}

\begin{document}
\title{Electric transitions of the charmed-strange mesons in a relativistic quark model}
\author{Shao-Feng Chen}
\author{Jing Liu}
\author{Hai-Qing Zhou}
\author{Dian-Yong Chen \footnote{Corresponding Author}}\email{chendy@seu.edu.cn}
\affiliation{
 School of Physics, Southeast University,  Nanjing 210094, China\\}
\date{\today}
\begin{abstract}
In the present work, we adopt a relativistic constituent quark model to depict the charmed strange meson spectroscopy, in which $D_{s0}(2317)$ and $D_{s1}(2460)$ are considered as the $1^3P_0$ and $1P_1^\prime$ charmed strange mesons, respectively. By using the wave function obtained from the relativistic quark model, we further investigate the electric transitions between charmed strange mesons. We find the long wave length approximation is reasonable for the charmed strange meson radiative decay by comparing the results with different approximations. The estimated partial widths are all safely under the upper limits of the experimental data. Moreover, we find the branching ratio of $D_{s1}(2536) \to D_s^\ast \gamma/D_s \gamma$ are large enough to be detected, which could be searched by further experiments in Belle II and LHCb. 
\end{abstract}
\maketitle
\section{Introduction}

Charmed-strange meson is one of important members of meson family. The ground $S-$wave states, $D_s$ and $D_s^\ast$, were first observed more than 40 years ago in the $e^+e^-$ annihilation process by DASP Collaboration~\cite{Brandelik:1977fg}\footnote{In Ref \cite{Brandelik:1977fg}, these two states named $F$ and $F^\ast$}. Later, in the $\bar \nu N$ collisions, a new state, $D_{s1}(2536)$, was observed in the $D_s^\ast \gamma$ invariant mass spectrum~\cite{Asratian:1987rb}, which could be a ground $P-$wave state. The second $P-$ wave states $D_{s2}^\ast(2537)$ was observed in the $D K$ and $D^\ast K$ modes in the $B$ meson decay processes by CLEO Collabortaion~\cite{Kubota:1994gn}. 

Nearly ten years later after the observation of $D_{s2}(2537)$, the rest two ground $P-$wave state candidates, $D_{s0}^\ast (2317)$ and $D_{s1}(2460)$, were discovered~\cite{Aubert:2003fg,Krokovny:2003zq}. The former one was first observed in the $D_s \pi^0$ invariant mass spectrum of $B$ decay process by BaBar Collaboration~\cite{Aubert:2003fg} and the later one was reported in a similar process but in the $D_s^\ast \pi^0$ invariant mass spectrum by Belle Collaboration~\cite{Krokovny:2003zq}. These two states are particular interesting since their observed masses are much lower than the quark model expectation~\cite{Godfrey:1985xj} and several tens MeV below the threshold of $DK$ and $D^\ast K$, respectively. Thus, these two states were ever considered as $DK$ and $D^\ast K$ molecular states due to their particular properties \cite{Xie:2010zza, Zhang:2006ix, Bicudo:2004dx, Faessler:2007gv, Faessler:2007us, Cleven:2014oka, Datta:2003re, Xiao:2016hoa}. However, considering the coupled channel effects and the fact that there are no additional states around quark model predicted masses, the authors in Refs.~\cite{Bracco:2005kt, Lutz:2008zz, Hwang:2004cd, Liu:2006jx, Lu:2006ry,  Liu:2013maa,Wang:2006mf,Fajfer:2015zma, Song:2015nia} assigned these two states as $P-$ wave charmed-strange mesons. In this case, the ground $P-$wave charmed strange mesons are established.

In 2006, the BaBar Collaboration reported two new charmed-strange meson in the $D K$ invariant mass spectrum of $B$ meson decay~\cite{Aubert:2006mh}. The narrow one is $D_{sJ}(2860)$ and the broader one is $D_{s1}^\ast(2700)$. The theoretical estimation indicate that the $D_{s1}^\ast (2700) $ could be a good candidate of $2^3S_1$ state \cite{Wang:2009as, Colangelo:2007ds, Zhang:2006yj, Song:2015nia}. In 2014, the LHCb Collaboration analyzed the $\bar{D} K$ invariant mass spectrum of $B_s^0 \to \bar{D} K^- \pi^+ $ process and find the structure around 2860 MeV announced by BaBar Collaboration consist of two particle with spin-1 and spin-3 \cite{Aaij:2014xza}, which were named as $D_{s1}^\ast (2860)$ and $D_{s3}^\ast(2860)$. As indicated in Refs.~\cite{Song:2014mha, Wang:2014jua,Godfrey:2014fga,Ke:2014ega}, these two states could be good candidates of $D-$wave charmed-strange mesons $1^3D_1$ and $1^3D_3$, respectively. To data, the observed heaviest charmed strange meson is $D_{sJ}(3040)$, which was discovered in $D^\ast K$ invariant mass spectrum of $B$ decay process by BaBar Collaboration~\cite{Aubert:2009ah}, which can be assigned as  $2P_1$ state as indicated in Refs.~\cite{Song:2015nia,Godfrey:2015dva}.

\begin{figure}
	\scalebox{0.48}{\includegraphics{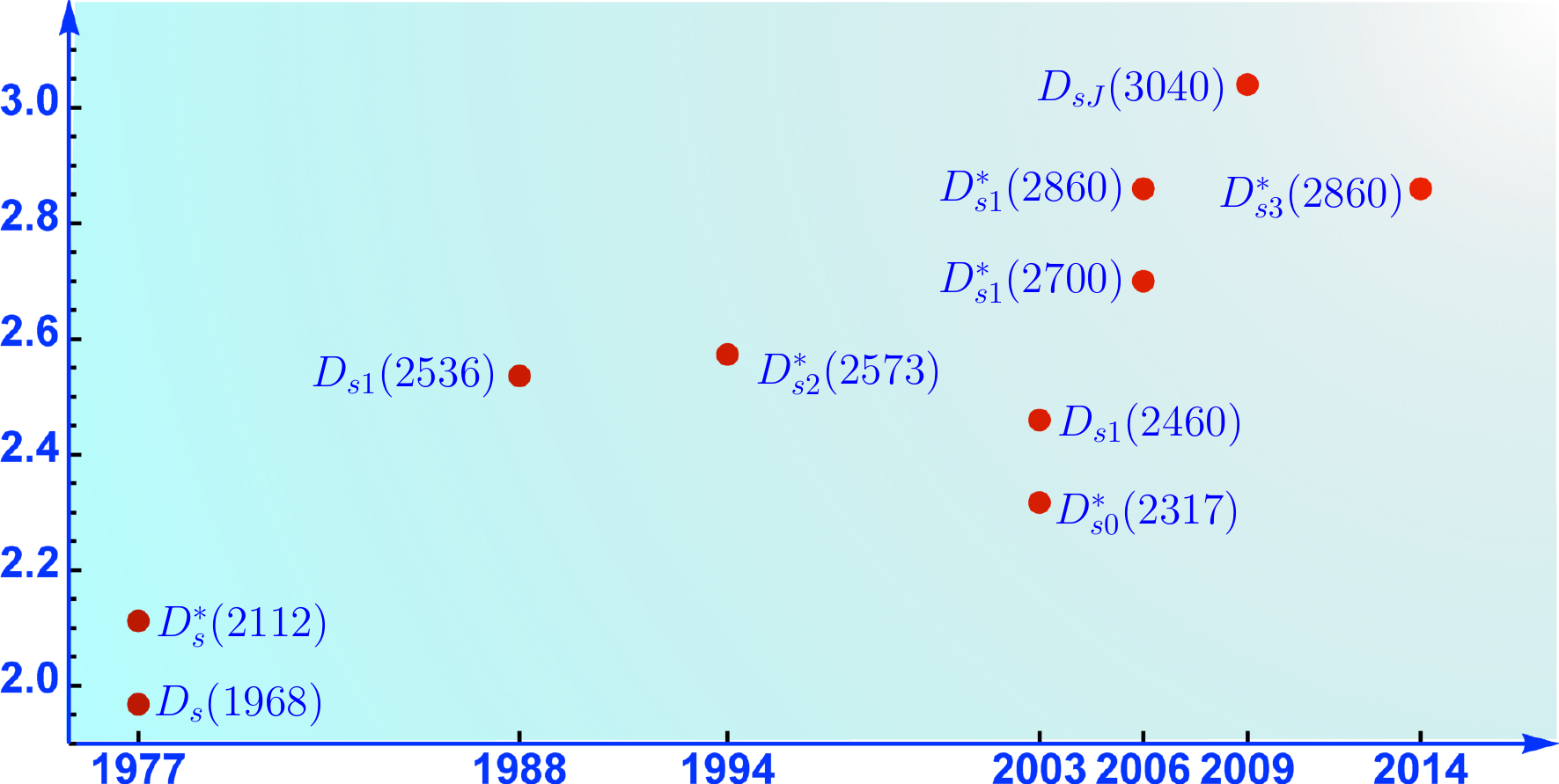}}
	\caption{The history of charmed-strange meson discovery~\cite{Brandelik:1977fg, Asratian:1987rb, Kubota:1994gn, Aubert:2003fg, Krokovny:2003zq, Aubert:2006mh, Aaij:2014xza}. Here, the masses of the charmed-mesons are taken from the Review of Particle Physics~\cite{Tanabashi:2018oca}. \label{Fig:Discovery}}
\end{figure}

In Fig.~\ref{Fig:Discovery}, we present the history of the observation of charmed-strange mesons, where we find most excited charmed-strange mesons were observed during the year of 2003-2014. Moreover, from the discovery history one can find most of the charmed-strange mesons are firstly observed in the bottom or bottom-strange meson decays. With the running of Belle II and LHCb, more excited charmed strange meson are expected to be discovered in the bottom or bottom-strange meson decays, which will make the charm-strange family abundant. 

Besides the observed resonance parameters, i.e., the mass and width, the decay behaviors of the observed states are also crucial to understand their inner structures. In particular, the electromagnetic transitions can be well described by Quantum electrodynamics in the quark level, which is unlike the  non-perturbative strong interactions in the hadron energy. Thus, the electromagnetic transitions could reflect the inner structure in a more comprehensive manner. On the experimental side, there are some experimental measurements for the radiative transitions between charmed mesons, the corresponding experimental information are collected in Table~\ref{Tab:Exp}. Thus the investigation of the radiative decays of charmed strange mesons are interesting and necessary. 

 \begin{table}
 \caption{Experimental information of the radiative transitions between charmed mesons. \label{Tab:Exp}}
\begin{tabular}{p{2.0cm}<{\centering} p{2.0cm}<{\centering} p{3.0cm}<{\centering}}
 	\toprule[1pt]
Initial  & Final & Experiments \cite{Tanabashi:2018oca}\\
 	\midrule[1pt]
 $D_s^\ast$ & $D_s$ & $(93.5\pm 0.7)\%$\\
 $D_{s0}(2317)$ & $D_s$ & $<5\%$\\
               & $D_s^\ast $ & $<6\%$\\
 $D_{s1}(2460)$ &  $D_s$ & $(18 \pm 4)\%$\\
             & $D_s^\ast $ & $<8\%$\\
           & $D_{s1}(2317)$ &$(3.7^{+5.0}_{-2.4})\%$\\
 $D_{s1}(2536)$ & $D_s^\ast $ & Possibly Seen\\
\bottomrule[1pt]
 	\end{tabular}
 \end{table}

In the present work, the charmed meson spectroscopy is depicted by a relativistic quark model~\cite{Liu:2013maa}, where the masses of the charmed strange mesons are well reproduced. With the wave functions estimated in the quark model, we estimate the electric transitions between the charmed-strange mesons, which could not only further test the relativistic quark model by comparing the theoretical estimations with the experimental measurements but also provide some useful predictions. 

This work is organized as follows, in Section~\ref{Sec:Spec}, we present a short review of the relativistic quark model, by which the mass spectroscopy of charmed strange mesons are estimated. In Section~\ref{Sec:Rad}, we present the formula of the electric transitions between charmed-strange mesons, and the numerical results and discussions are presented in Section~\ref{Sec:Num}. A short summary is given in Section~\ref{Sec:Sum}.

\section{Review of MASS SPECTROSCOPY of charmed strange mesons  }
\label{Sec:Spec}

Relativistic quark model is usually adopted to   depict the mass spectroscopy of hadrons since the non-perturbative properties of QCD in hadron energy. Such kind of quark model was proposed to investigate the  meson spectroscopies systematically in 1985 by Godfrey and Isgur \cite{Godfrey:1985xj}. In this model, the mass spectroscopy and wave functions of the mesons can be determined by solving the relativistic Schr\"odinger equation, where the spin independent Hamiltonian can be,
\begin{eqnarray}
	H_0 =\sqrt{p^2+m_1^2}+\sqrt{p^2+m_2^2}+V(r)
\end{eqnarray}
where $r$, $p$ are the coordinates and the momentum of quark in the center-of-mass frame, respectively.  $m_1$ and $m_2$ are the masses of the quark and the antiquark, respectively. $V(r)$ is the effective spin-independent potential between the quark and the antiquark, including a Coulomb term and a linear confining term \cite{Godfrey:1985xj}, which is
\begin{eqnarray}
V(r)=-\frac{4 \alpha_{s}\left( r \right)}{3 r}+b r+c
\end{eqnarray}
As for the spin-dependent part $H'$, it includes the spin-spin interaction and spin-orbital interactions, which is,
\begin{eqnarray}
H'=H_{SS}+H_{SL}
\end{eqnarray}
and the concrete form of spin-spin and spin-orbital interactions are
\begin{eqnarray}
H_{SS}&=&f(r)\vec{s_{1}}\cdot\vec{s_{2}}+g(r)\left(\frac{3\vec{s_{1}}\cdot\vec{r}\vec{s_{2}}\cdot\vec{r}}{r^{2}}-\vec{s_{1}}\cdot\vec{s_{2}}\right) \nonumber\\
H_{SL}&=&h_{1}(r)\vec{s_{1}}\cdot\vec{L}+h_{2}(r)\vec{s_{2}}\cdot\vec{L}	
\end{eqnarray}
where the functions $f(r),g(r),h_{1}(r),h_{2}(r)$ can be found in Ref.~\cite{Liu:2013maa}. With these spin dependent terms, the $S$-$D$ mixings and spin-singlet and spin-triplet mixings have been included. In this model, the mass spectroscopy of the charmed strange mesons can be well reproduced, thus, in the present work, we adopt the same model parameters as those in Ref.~\cite{Liu:2013maa} to investigate the electric radiative decays of the charmed strange mesons. Before the estimations of the radiative decays, we present the mass spectroscopy of the charmed-strange mesons in Table \ref{Tab:Mass}, where the theoretical estimations from Ref.~\cite{Godfrey:2015dva,Devlani:2011zz}  and experimental data~\cite{Tanabashi:2018oca} are also listed for comparison. The theoretical estimated mass of $1^3P_0$ and $1P_1$ states are 2317 and 2425 MeV, respectively, which are more consistent with the experimental measurements comparing to other works~\cite{Godfrey:2015dva, Devlani:2011zz}.

\begin{table}[htb]
\caption{spectrum of the charmed-strange mesons in unit of MeV. For comparison, we also list the theoretical estimations in Refs.~\cite{Godfrey:2015dva,Devlani:2011zz} and experimental measurements \cite{Tanabashi:2018oca}}\label{Tab:Mass}
\centering
\begin{tabular}{c|p{1.0cm}<{\centering} p{1.25cm}<{\centering} p{1.25cm}<{\centering} p{1.25cm}<{\centering} p{2cm}<{\centering}}
\toprule[1pt]
\multicolumn{2}{c}{States} &  Present & Ref~\cite{Godfrey:2015dva} & Ref~\cite{Devlani:2011zz} & PDG~\cite{Tanabashi:2018oca} \\
\midrule[1pt]
\multirow{6}{0.5cm}{\rotatebox{90}{S Wave} }
 & $1^1S_0$ &	$1964$ & $1979$ & $1970$ & $1968.34\pm 0.07$ \\
 & $1^3S_1$ & $2102$ & $2129$ & $2117$ & $2112.2\pm 0.4$\\	
 & $2^1S_0$ &  $2557$ & $ 2673$ & $2684$ &\\
 & $2^3S_1$ &  $2680$ & $2732$ & $2723$  &   $2708.3^{+4.0}_{-3.4}$\\
 & $3^1S_0$ &  $2999$ &$ 3154   $ & $3158$ &\\	
 & $3^3S_1$ &  $3105$ & $3193$ &  $3180$ &\\	
\midrule[1pt]
\multirow{6}{0.5cm}{\rotatebox{90}{P Wave} }
 & $1^3P_0$ &  $2317$ &  $2484$ &  $2444$  &  $2317.8\pm 0.5$ \\	
 & $1 P_1$ &   $2425$ &  $2549$  &  $2530$  &  $2459.5\pm 0.6$\\	
 & $1 P_1^\prime$ &  $2510$ &  $2556$  &  $2540$  &  $2535.11\pm 0.06$\\	
 & $1^3P_2$ &  $2548$  &    $2592$  &  $2566$  &  $2569.1\pm 0.8$\\	
 & $2^3P_0$ &  $2700$  &$3005$  &  $2947$ \\	
 & $2 P_1^\prime$ &  $2876$  &  $3018$  &  $3019$\\	
 & $2 P_1$ &  $2965$ &  $3038$  &  $3023$ & $3044\pm 8^{+30}_{-5}$\\	
 & $2^3P_2$ &  $3019$ &  $3048$  &  $3048$\\	
\midrule[1pt]
\multirow{6}{0.5cm}{\rotatebox{90}{D Wave} }
 & $1^3D_1$ &  $2771$ &  $2899$  &  $2873$  &  $2859\pm 27$  \\
 & $1 D_2$ &  $2800$ &  $2900$  &  $2816$\\
 & $1 D_2^\prime$ &  $2826$  &  $2926$  &  $2896$\\
 & $1^3D_3$ &  $2816$  &  $2917$ &  $2834$  &  $2860.1\pm 7$\\
 & $2^3D_1$ &  $3138$  &  $3306$ &  $3292$  \\
 & $2 D_2$ & $3191$  &  $3323$ &  $3312$\\
 & $2 D_2^\prime$ &  $3186$  &  $3298$  &  $3248$\\
 & $2^3D_3$ &  $3214$  &  $3311$  &  $3263$\\
 \bottomrule[1pt]
\end{tabular}
\end{table}

\section{Electric transitions of the charmed strange mesons}
\label{Sec:Rad}

In the quark level, the quark-photon electromagnetic interaction can be written as
\begin{eqnarray}
H_{e}=-\sum_{j}e_{j}\bar{\psi}_{j}\gamma_{\mu}\mathcal{A}^{\mu}(\textbf{k},\textbf{r})\psi_{j}
\label{Eq:He}
\end{eqnarray}
where $\psi_{j}$ and $e_{j}$ represent the $j-th$ quark fields and its charges in the charmed-strange meson, respectively. The $\textbf{k}$ is three momentum of the emitted photon. After performing some algebra estimation as shown in Appendix \ref{Sec:App}, the amplitude of the electromagnetic transition can be expressed,
\begin{eqnarray}
\left\langle f\left| H_e \right|i \right\rangle &=& \left\langle f\left| \alpha \cdot \epsilon e^{i \mathbf{k}\cdot \mathbf{r}_j} \right| i\right\rangle\nonumber\\
&=& -i\omega \left\langle f\left| r_j \cdot \epsilon e^{i \mathbf{k}\cdot \mathbf{r}_j} (1-\alpha \cdot \hat{k})\right|i \right\rangle 
\label{Eq:EM}
\end{eqnarray}
where $|i\rangle$ and $|f\rangle$ are the wave functions of initial and final states, respectively. $\omega$ is the energy of the emitted photon.

In the present work, we mainly focus on the electric transition processes, and the helicity amplitude is
\begin{eqnarray}
\mathcal{A}^{E}_{\lambda}=-i\sqrt{\frac{\omega}{2}}\langle f|\sum_{j}e_{j}\textbf{r}_{j}\cdot\boldsymbol{\epsilon}e^{-i \mathbf{k} \cdot \mathbf{r}_{j}}| i\rangle
\end{eqnarray}
where the initial and final hadron wave functions can be estimated by the relativistic quark model. In the estimation, we can choose the photon momentum direction along the $z$ axis, i.e., $\textbf{k}=k\hat{\textbf{z}}$, and the photon polarization vector is in right-hand form, which is $\boldsymbol{\epsilon}=-(1,i,0)/\sqrt{2}$. In this case, $e^{-i \mathbf{k} \cdot \mathbf{r}_{j}}$ can be expanded as,
\begin{eqnarray}
e^{-i \mathbf{k} \cdot \mathbf{r}_{j}}=\sum _{l}\sqrt{4\pi (2l+1)}(-i)^{l}j_{l}(kr_j)Y_{l0}(\Omega),
\end{eqnarray}
then the helicity amplitude for the angular momentum $l$ can be~\cite{Deng:2016stx},
\begin{eqnarray}
\mathcal{A}^{E}_{l,\lambda}=\sqrt{\frac{\omega}{2}}\langle f|\sum_{j}(-i)^{l}\sqrt{\frac{2\pi l(l+1)}{2l+1}}e_{j}j_{l+1}(kr_j)r_{j}Y_{l1}| i\rangle\\ \nonumber
+ \sqrt{\frac{\omega}{2}}\langle f|\sum_{j}(-i)^{l}\sqrt{\frac{2\pi l(l+1)}{2l+1}}e_{j}j_{l-1}(kr_j)r_{j}Y_{l1}| i\rangle,
\end{eqnarray}
and then the decay width of the electric transition between $Q\bar{q}$ can be estimated as
\begin{eqnarray}
\Gamma(A\rightarrow B \gamma)&=&\sum_{k=0,2} \frac{4\alpha}{3}\omega^3 C_{fi}\delta_{SS'}\delta_{LL'\pm1}\nonumber\\
&&\Big|\Big\langle n'^{2S'+1}L'_{J'}\Big|\frac{e_q m_Q r}{m_q+m_Q}j_k \left(\frac{m_2 \omega r}{m_q+m_Q}\right) \nonumber\\
&&-\frac{e_{\bar{Q}}m_{q}}{m_{q}+m_{Q}}j_k \left(\frac{m_1\omega r}{m_{q}+m_{Q}}\right)\Big|n^{2S+1}L_{J}\Big\rangle\Big|^{2}
\label{Eq:ModeI}
\end{eqnarray}
where $|n'^{2S'+1}L'_{J'} \rangle$ and $|n^{2S+1}L_{J} \rangle$ represent the final and initial states, respectively. $C_{fi}$ is a coefficient related to involved states, which is 
\begin{eqnarray}
C_{fi}= \mathrm{max}(L_A,L_B)(2J_B+1)\left\{
\begin{array}{ccc}
 	L_B & J_B & S\\
	J_A & L_A & 1
\end{array}	
\right\}
\end{eqnarray}
Considering the lowest order of the electric transition, the terms related to $j_2(kr)$ can be ignored, then the electric transition width can be
\begin{eqnarray}
\Gamma(A\rightarrow B \gamma)&=& \frac{4\alpha}{3}\omega^3 C_{fi}\delta_{SS'}\delta_{LL'\pm1}\nonumber\\
&&\Big|\Big\langle n'^{2S'+1}L'_{J'}\Big|\frac{e_q m_Q r}{m_q+m_Q}j_0 \left(\frac{m_2 \omega r}{m_q+m_Q}\right) \nonumber\\
&&-\frac{e_{\bar{Q}}m_{q}}{m_{q}+m_{Q}}j_0 \left(\frac{m_1\omega r}{m_{q}+m_{Q}}\right)\Big|n^{2S+1}L_{J}\Big\rangle\Big|^{2}
\label{Eq:ModeII}
\end{eqnarray}
In the literatures, the zeroth order spherical Bessel function $j_0(kr)$ is usually expanded as $j_0(kr)=1+\mathcal{O}(x^2)$, keeping the lowest order, one can get the partial width as 
\begin{eqnarray}
\Gamma(A\rightarrow B \gamma)=\frac{4\alpha}{3}e_{M}^{2}\omega^{3}C_{fi}\delta_{SS'}\delta_{LL'\pm1}|\langle n'^{2S'+1}L'_{J'}|r|n^{2S+1}L_{J}\rangle|^{2}\nonumber\\
\label{Eq:ModeIII}
\end{eqnarray}
where $e_M=(e_{\bar{Q}} mq-e_q m_{Q})/(m_q+m_Q)$. The approximation in above formula  corresponds to the long wave length approximation, where $e^{i \textbf{k}\cdot \textbf{r}} \sim 1$.
 
\begin{table*}[htb]
\centering
\caption{Electric transition widths for  $S \to P\gamma $ processes, where $P$ and $S$ are $P-$ and $S-$wave charmed-strange mesons, respectively. For comparison, we also present the theoretical estimations from Refs.~\cite{Radford:2009bs,Green:2016occ,Close:2005se,Goity:2000dk}. The results of Ref.~\cite{Close:2005se} are estimated with the mixing angle $\theta_{1P}=-38^\circ$ \cite{Godfrey:1986wj, Godfrey:2003kg} \label{Tab:S2P} }
\begin{tabular}{ c c c c c c c c ccc }
\toprule[1pt]
 \multirow{2}{1.2cm}{~Initial} & \multirow{2}{1.2cm}{~Final}& \multicolumn{9}{c}{ Decay Width (keV)} \\
\cline{3-11}
  &  & Mode I & Mode II & Mode III & Ref.~\cite{Radford:2009bs} &  Ref.~\cite{Green:2016occ}  &Ref.~\cite{Close:2005se}    & \multicolumn{3}{c}{Ref.~\cite{Goity:2000dk}}\\
\midrule[1pt]
$D_{s}(2^{1}S_{0})$ & $D_{s1}(1P_{1})$  &  $0.07$  &  $0.07$  &  $0.07$  &  $0.01$  &  $0.05$ &  $3.35$ & $3.3\pm 0.6$ & $4.0 \pm 0.7$ & $3.8 \pm 0.7$\\
  & $D_{s1}(1P_{1}^\prime)$  &  $0.18$  &  $0.18$  &  $0.18$ & & & $0.57$  &  $4.6 \pm 1.2$ & $4.3 \pm 1.1$ & $4.3 \pm 1.1$\\
$D_{s}^{\ast}(2^{3}S_{1})$ &  $D_{s0}^{\ast}(1^{3}P_{0})$  &  $3.32$  &  $3.20$ & $3.20$  &  $6.76$  &  $8.77$  &  &  $2.4 \pm 0.0$  &  $2.6 \pm 0.1$  &  $2.5 \pm 0.1$ \\
  &  $D_{s1}(1P_{1})$  &  $1.22$  &  $1.20$ & $1.20$  &  $2.8$  &  $4.25$  &    &  $4.0 \pm 0.2$  &  $4.9 \pm 0.2$  &  $4.8 \pm 0.2$\\
  &  $D_{s1}(1P_{1}^\prime)$  &  $0.30$  &  $0.30$ & $0.30$  &  $0.24$  &  $0.41$ &    &  $2.1\pm 0.3$  &  $2.1 \pm 0.4$  &  $2.1 \pm 0.3$ \\
  &  $D_{s2}^{\ast}(1^{3}P_{2})$  &  $1.21$  &  $1.21$ & $1.21$  &  $0.35$  &  $0.71$ &    &  $8.1 \pm 1.1$  &  $7.5 \pm 1.1$  &  $7.6 \pm 1.1$ \\
  \midrule[1pt]
$D_{s}(3^{1}S_{0})$ & $D_{s1}(1P_{1})$  &  $1.58$  &  $1.60$  &  $1.60$  \\
  & $D_{s1}(1P_{1}^\prime)$  &  $1.10$  &  $0.76$  &  $0.77$ \\
  & $D_{s1}(2P_{1})$  &  $0.77$  &  $0.76$  &  $0.76$  \\
  & $D_{s1}(2P_{1}^\prime)$  &  $0.05$  &  $0.05$  &  $0.05$ \\
$D_{s}^{\ast}(3^{3}S_{1})$  &  $D_{s0}^{\ast}(1^{3}P_{0})$  &  $1.74$  &  $1.16$ & $1.17$  \\
  &  $D_{s1}(1P_{1})$  &  $2.00$  &  $2.06$ & $2.06$  \\
  &  $D_{s1}(1P_{1}^\prime)$  &  $0.39$  &  $0.28$ & $0.28$  \\
  &  $D_{s2}^{\ast}(1^{3}P_{2})$  &  $0.99$  &  $0.65$ & $0.66$  \\
  &  $D_{s0}^{\ast}(2^{3}P_{0})$  &  $14.06$  &  $13.04$ & $13.05$  \\
  &  $D_{s1}(2P_{1})$  &  $3.91$  &  $3.80$ & $3.80$  \\
  &  $D_{s1}(2P_{1}^\prime)$  &  $0.34$  &  $0.33$ & $0.33$  \\
  &  $D_{s2}^{\ast}(2^{3}P_{2})$  &  $0.69$  &  $0.69$ & $0.69$  \\
\bottomrule[1pt]
\end{tabular}
\end{table*}

\begin{table*}[htb]
\centering
\caption{The same as Table~\ref{Tab:S2P} but for $P\to S \gamma$ processes.For comparison, we also present the theoretical estimations from Refs.~\cite{Radford:2009bs,Green:2016occ,Close:2005se,Goity:2000dk,Korner:1992pz,Godfrey:2005ww}. The results of Ref.~\cite{Close:2005se} are estimated with the mixing angle $\theta_{1P}=-38^\circ$ \cite{Godfrey:1986wj, Godfrey:2003kg}  \label{Tab:P2S} }
\begin{tabular}{ c c c c c c c c c c ccc}
\toprule[1pt]
 \multirow{2}{1.2cm}{~Initial} & \multirow{2}{1.2cm}{~Final}& \multicolumn{11}{c}{ Decay Width (keV)} \\
\cline{3-13}
  &  & Mode I & Mode II & Mode III & Ref.~\cite{Radford:2009bs} &  Ref.~\cite{Green:2016occ} &  Ref.~\cite{Korner:1992pz}  &  Ref.~\cite{Godfrey:2005ww}  &   Ref.~\cite{Close:2005se} &
  \multicolumn{3}{c}{Ref.~\cite{Goity:2000dk}}\\
\midrule[1pt]
$D_{s0}^{\ast}(1^{3}P_{0})$  &  $D_{s}^{\ast}(1^{3}S_{1})$ & $2.07$ & $2.06$ &  $2.06$  &  $4.92$  &  $5.46$  &  &  $1.9$   & $1.0$  &$24.9\pm 1.9$ &  $14.5\pm 0.9$ &  $16.2 \pm 1.0$ \\
$D_{s1}(1P_{1}^\prime)$  & $D_{s}(1^{1}S_{0})$ & $3.61$  &  $3.53$  &  $3.53$  &  $12.8$  &  $13.2$  &  & $15.0$  & $4.02$ &  $25.2 \pm 0.5$  &$31.1 \pm 0.8$  & $30.0 \pm 0.7$  \\
  & $D_{s}^{\ast}(1^{3}S_{1})$ & $4.79$  &  $4.74$  &  $4.74$ &  $15.5$  &  $17.4$  &  & $5.6$ & $4.41$  &  $14.6 \pm 0.2$  &  $22.8  \pm 1.2 $  & $21.0  \pm 1.0$\\
$D_{s1}(1P_{1})$  & $D_{s}(1^{1}S_{0})$ & $18.85$  &  $18.18$  &  $18.18$  &  $54.5$  &  $61.2$  & $1.6 \pm 2.3$  &  $6.2$  & $4.53$  & $17.2 \pm 0.7$  &  $10.3 \pm 0.6$  &  $11.4 \pm 0.6$\\
  & $D_{s}^{\ast}(1^{3}S_{1})$ & $3.02$  &  $2.96$  &  $2.96$   &  $8.90$  &  $9.21$  &  $0.4 \pm 1.0$  &   $5.5$  & $1.59$   &  $25.1 \pm 1.4$   &  $14.0 \pm 0.8 $  &  $15.8  \pm 0.9$\\
$D_{s2}^{\ast}(1^{3}P_{2})$  & $D_{s}^{\ast}(1^{3}S_{1})$ & $15.66$  &  $15.23$  &  $15.23$  &  $44.1$  &  $49.6$   &  $1.4 \pm 2.0$ &  $19.0$  & $8.8$  &  $41.5 \pm 0.0$  &  $55.9^{+0.9}_{-0.6}$  &  $53.0^{+0.4}_{-0.5}$ \\
\midrule[1pt]
$D_{s0}^{\ast}(2^{3}P_{0})$  &  $D_{s}^{\ast}(1^{3}S_{1})$ & $0.03$ & $0.03$ &  $0.03$ \\
  &  $D_{s}^{\ast}(2^{3}S_{1})$ & $0.004$ & $0.004$ &  $0.004$ \\
$D_{s1}(2P_{1}^\prime)$  & $D_{s}(1^{1}S_{0})$ & $0.91$  &  $0.55$  &  $0.56$  \\
  & $D_{s}^{\ast}(1^{3}S_{1})$ & $2.45$  &  $1.78$  &  $1.79$ \\
  & $D_{s}(2^{1}S_{0})$ & $4.19$  &  $4.04$  &  $4.04$  \\
  & $D_{s}^{\ast}(2^{3}S_{1})$ & $3.15$  &  $3.11$  &  $3.11$ \\
$D_{s1}(2P_{1})$  & $D_{s}(1^{1}S_{0})$ & $0.46$  &  $1.05$  &  $1.07$  \\
  & $D_{s}^{\ast}(1^{3}S_{1})$ & $0.05$  &  $0.13$  &  $0.13$ \\
  & $D_{s}(2^{1}S_{0})$ & $16.89$  &  $15.90$  &  $15.90$  \\
  & $D_{s}^{\ast}(2^{3}S_{1})$ & $2.45$  &  $2.38$  &  $2.38$ \\
$D_{s2}^{\ast}(2^{3}P_{2})$  & $D_{s}^{\ast}(1^{3}S_{1})$ & $1.71$  &  $2.53$  &  $2.54$  \\
  & $D_{s}^{\ast}(2^{3}S_{1})$ & $12.89$  &  $12.23$  &  $12.23$  \\
\bottomrule[1pt]
\end{tabular}
\end{table*}

\section{Numerical Results and Discussions}
\label{Sec:Num}
As indicated in the last section, the partial widths of electric transition can be estimated with different approximations, hereafter, we use Mode I, Mode II and Mode III to refer the estimations with Eqs.(\ref{Eq:ModeI}), (\ref{Eq:ModeII}) and (\ref{Eq:ModeIII}), respectively and further check the reliability of different approximations. With the wave functions estimated from the relativistic quark model and the formula in above section, we can get the partial widths of the electric transitions, which are listed in Tables~\ref{Tab:S2P}-\ref{Tab:D2P}.

In Table~\ref{Tab:S2P}, we present the electric transitions for $P\to S\gamma$ processes, where $P$ and $S$ indicate the $P-$ and $S-$ wave charmed-strange mesons, respectively. In addition, we also listed the theoretical results from other groups~\cite{Radford:2009bs,Green:2016occ, Goity:2000dk, Close:2005se} for comparison. From the table, one can find the estimation from different approximations are almost the same, which indicates that the approximation from Mode I to Mode III are still reliable and long wave length approximation in the considered electric transitions of charmed-strange mesons is reasonable. Our estimation indicates that most of our results are of the same order as those in Refs.~\cite{Radford:2009bs,Green:2016occ, Goity:2000dk, Close:2005se}. In particular, we find the partial widths of $D_s(2^1S_0) \to D_{s1}(1P_1) \gamma$ and $D_s(2^1S_0) \to D_{s1}(1P_1^\prime) \gamma$ from different literatures are very different. Our estimation shows that the partial width of $D_s(2^1S_0) \to D_{s1}(1P_1) \gamma$ is 0.07 keV, which is of same order as those in Ref.~\cite{Radford:2009bs,Green:2016occ}, but the estimation in Ref.~\cite{Goity:2000dk, Close:2005se} are about two order larger than the present estimation.  As for $D_s(2^1S_0) \to D_{s1}(1P_1^\prime) \gamma$, our estimation is of the same order as the one in Ref.~\cite{Close:2005se}, but much smaller than those in Ref.~\cite{Goity:2000dk}. The estimations in the present work and in Refs. \cite{Radford:2009bs, Green:2016occ, Close:2005se} are all based on relativistic quark model. But it should be notice that the estimated mass spectroscopy in Refs.~\cite{Radford:2009bs, Green:2016occ} are similar to the present one, where the masses of $D_{s0}(2317)$ and $D_{s1}(2460)$ were well reproduced. Thus the meson  wave functions should be similar and so do the electric transition widths. As for Ref.~\cite{Goity:2000dk}, the estimated mass spectroscopy are much different with the present one and the estimated masses of $D_{s0}(2317)$ and $D_{s1}(2460)$ are far above the measured values, then the meson wave functions and electric transition widths are much different. As for Ref. \cite{Close:2005se}, the estimations are based on heavy quark limit, which should be more   reliable for bottom mesons.

 As for the radiative decay of $3S$ states, we find the that $\Gamma(D_{s}^\ast(3^3S_1) \to D_{s0}^\ast(2^3P_0) \gamma)=(13 \sim 14)$ keV. As for $D_{s}^\ast(3^3S_1)$, it is far above the threshold of $DK$, and it dominantly decay into a charmed meson and a strange meson, and its total width are estimated to be around  $100$ MeV~\cite{Song:2015nia}, and with such a large width, the branching ratio of $D_{s}^\ast(3^3S_1) \to D_{s0}^\ast(2^3P_0) \gamma$ is of order $10^{-4}$.

In Table \ref{Tab:P2S}, we present our estimated widths for $P \to S\gamma$. Our estimation indicates that the partial widths for $1P \to 1S \gamma$ vary from several keV to 10 keV, which is consistent with those in the previous literatures~\cite{Radford:2009bs,Green:2016occ, Korner:1992pz, Godfrey:2005ww, Close:2005se, Goity:2000dk}. The partial width of $D_{s0}(1^3P_0) \to D_s^\ast \gamma$ is estimated to be around 2 keV. The measured upper limits of $\Gamma_{D_{s0}(1^3P_0)}$ and  $B(D_{s0}(1^3P_0) \to D_s^\ast \gamma)$ are $3.5$ MeV and $6\%$, respectively. Thus, the upper limit of the partial width of $D_{s0}(1^3P_0) \to D_s^\ast \gamma$ is 210 keV, which indicates our estimation is safely under the upper limit of the experimental values. 

As for $D_{s1}^\prime (2460)$, the widths of $D_s \gamma$ and $D_s^\ast \gamma$ modes are 3.61 and 4.79 keV, respectively, which are both safely under the upper limits of the experimental values. Moreover, from our estimation, we find that the partial width of $D_s^\ast \gamma$ mode is a bit larger than the one of $D_s \gamma$, which is similar to those in Refs.~\cite{Radford:2009bs,Green:2016occ, Goity:2000dk,Close:2005se}, but different with the experimental measurements, which are $B(D_{s1}^\prime(2460) \to D_s \gamma)=(18 \pm 4)\%$ and $B(D_{s1}^\prime(2460) \to D_s^\ast \gamma)<8\%$. It should be notice that the $D_{s1}^\prime(2460)$ state has the components with both $S=0$ (corresponding to $^1P_1$ state) and $S=1$ (corresponding to $^3P_1$ state), while in the electric transitions, the spin of the initial and final states should be the same, thus, the electric transitions involves $D_{s1}(nP_1^\prime)$ and $D_{s1}(nP_1)$ states are sensitive to the spin singlet and triplet mixing. 

As for $D_{s1}(1P_1)$ state, our estimation indicates that the partial widths of $D_s \gamma$ and $D_s^\ast \gamma$ are 18.85 and 3.02 keV, respectively. The width of $D_{s1}(1P_1)$ is measured to be $(0.92\pm 0.05)$ MeV, then the branching ratios of $D_{s1}(1P_1) \to D_s \gamma $ and $D_s^\ast \gamma$ can be $2.0\%$ and $3.3 \times 10^{-3}$, which should be large enough to be detected. On the experimental side, there may be some experimental hint of $D_{s1}(1P_1) \to D_s^\ast \gamma$ process. As for $D_{s2}(1^3P_2)$, we find the partial width of $D_{s2}(1^3P_2) \to D_s^\ast \gamma$ could reach up to 15.66 keV, which indicates the branching ratio is  about $9 \times 10^{-4}$.  As for $2P$ states, we find the partial widths of  $D_{s1}(2P_1) \to D_{s}(2^1S_0) \gamma$ and $D_{s2}^\ast(2^3P_2) \to D_{s}^\ast(2^3S_1) \gamma$ are more than 10 keV. As shown in Ref.~\cite{Song:2015nia}, the total widths of $D_{s1}(2P_1)$ and $D_{s2}(2^3P_2)$ are estimated to be $285.3$ and $86.25$ MeV, respectively. Thus, the branching ratios of $D_{s1}(2P_1) \to D_{s}(2^1S_0) \gamma$ and $D_{s2}^\ast(2^3P_2) \to D_{s}^\ast(2^3S_1) \gamma$ are of order $10^{-5}$ and $10^{-4}$, respectively. 

Our estimation for $P \to D \gamma$ and $D\to P \gamma$ process are listed in Table \ref{Tab:P2D} and \ref{Tab:D2P}.  As for $P \to D\gamma$ processes, the largest one is $D_{s2}^\ast (2^3P_2) \to D_{s3}(1^3D_3)$, which are 3.22 keV. As for $D \to P \gamma$ processes, the partial widths of $D_s^\ast (1^3D_1) \to D_{s0}^\ast(1^3P_0) \gamma$,  $D_s^\ast (1^3D_3) \to D_{s0}^\ast(1^3P_2) \gamma$ and $D_{s2}(2D_2) \to D_{s1}(2P_1^\prime) \gamma $ processes are greater than 10 keV. These highly excited states are far above the threshold of $DK$ and $D^\ast K$, and they dominantly decay into a charmed meson and a strange meson, their width should be of order 100 MeV. Thus the branching ratios of these radiative decays should be of order of $10^{-4}$.

\begin{table}[htb]
\centering 
\caption{Electric transition width for $P \to D\gamma $ processes, where $P$ and $D$ are $P-$ and $D-$wave charmed-strange mesons, respectively.  \label{Tab:P2D} }
\begin{tabular}{ ccccccc}
\toprule[1pt]
\multirow{2}{1.25cm}{~Initial} & \multirow{2}{1.25cm}{~Final} &  & Decay width (keV) \\
\cline{3-5}
  &  & Mode I & Mode II & Mode III &   &  \\
\midrule[1pt]
$D_{s1}(2P_{1}^\prime)$  & $D_{s}^{\ast}(1^{3}D_{1})$ & $0.11$  &  $0.11$  &  $0.11$  \\
  & $D_{s2}(1D_{2})$ & $0.15$  &  $0.15$  &  $0.15$  \\
  & $D_{s2}(1D_{2}^\prime)$ & $0.04$  &  $0.04$  &  $0.04$  \\
$D_{s1}(2P_{1})$  & $D_{s}^{\ast}(1^{3}D_{1})$ & $0.12$  &  $0.12$  &  $0.12$  \\
  & $D_{s2}(1D_{2})$ & $1.11$  &  $1.11$  &  $1.11$  \\
  & $D_{s2}(1D_{2}^\prime)$ & $0.45$  &  $0.44$  &  $0.44$  \\
$D_{s2}^{\ast}(2^{3}P_{2})$ & $D_{s}^{\ast}(1^{3}D_{1})$ & $0.31$  &  $0.30$  &  $0.30$  \\
  & $D_{s2}(1D_{2})$ & $0.25$  &  $0.24$  &  $0.24$  \\
  & $D_{s2}(1D_{2}^\prime)$ & $0.13$  &  $0.13$  &  $0.13$  \\
  & $D_{s3}(1^{3}D_{3})$ & $3.22$  &  $3.15$  &  $3.15$  \\
\bottomrule[1pt]
\end{tabular}
\end{table}

\begin{table}[htb]
\centering 
\caption{The same as Table~\ref{Tab:P2D} but for $D\to P \gamma$ process.\label{Tab:D2P} }
\begin{tabular}{ ccccccc}
\toprule[1pt]
\multirow{2}{1.25cm}{~Initial} & \multirow{2}{1.25cm}{~Final} &  & Decay width (keV) \\
\cline{3-5}
  &  & Mode I & Mode II & Mode III &   &  \\
\midrule[1pt]
$D_{s}^{\ast}(1^{3}D_{1})$  &  $D_{s0}^{\ast}(1^{3}P_{0})$  & $21.26$  &  $20.49$  &  $20.49$  \\
  &  $D_{s1}(1P_{1}^\prime)$  & $4.33$  &  $4.25$  &  $4.25$  \\
  &  $D_{s1}(1P_{1})$  & $0.87$  &  $0.86$  &  $0.86$ \\
    &  $D_{s2}^{\ast}(1^{3}P_{2})$  & $0.27$  &  $0.26$  &  $0.26$  \\
  &  $D_{s0}^{\ast}(2^{3}P_{0})$  & $0.03$  &  $0.03$  &  $0.03$ \\
$D_{s2}(1D_{2}^\prime)$  &  $D_{s1}(1P_{1}^\prime)$  & $6.26$  &  $6.12$  &  $6.12$ \\
  &  $D_{s1}(1P_{1})$  & $6.33$  &  $6.21$  &  $6.21$ \\
  &  $D_{s2}^{\ast}(1^{3}P_{2})$  & $1.10$  &  $1.09$  &  $1.09$ \\
$D_{s2}(1D_{2})$  &  $D_{s1}(1P_{1}^\prime)$  & $8.59$  &  $8.37$  &  $8.37$  \\
  &  $D_{s1}(1P_{1})$  & $7.10$  &  $6.95$  &  $6.95$ \\
  &  $D_{s2}^{\ast}(1^{3}P_{2})$  & $1.72$  &  $1.69$  &  $1.69$ \\
$D_{s3}(1^{3}D_{3})$ &  $D_{s2}^{\ast}(1^{3}P_{2})$ &  $11.77$  &  $11.56$  & $11.56$  \\
\midrule[1pt]
$D_{s}^{\ast}(2^{3}D_{1})$  &  $D_{s0}^{\ast}(1^{3}P_{0})$  & $3.96$  &  $2.33$  &  $2.37$  \\
  &  $D_{s1}(1P_{1})$  & $0.80$  &  $0.88$  &  $0.88$  \\
  &  $D_{s1}(1P_{1}^\prime)$  & $0.71$  &  $0.54$  &  $0.54$ \\
  &  $D_{s2}^{\ast}(1^{3}P_{2})$  & $0.32$  &  $0.23$  &  $0.23$ \\
  &  $D_{s0}^{\ast}(2^{3}P_{0})$  & $32.75$  &  $30.25$  &  $30.26$  \\
  &  $D_{s1}(2P_{1}^\prime)$  & $4.99$  &  $4.83$  &  $4.83$  \\
  &  $D_{s1}(2P_{1})$  & $0.52$  &  $0.52$  &  $0.52$ \\
  &  $D_{s2}^{\ast}(2^{3}P_{2})$  & $0.69$  &  $0.69$  &  $0.69$ \\
$D_{s2}(2D_{2}^\prime)$  &  $D_{s1}(1P_{1}^\prime)$  & $2.73$  &  $2.81$  &  $2.81$ \\
  &  $D_{s1}(1P_{1})$  & $0.54$  &  $0.19$  &  $0.21$ \\
  &  $D_{s2}^{\ast}(1^{3}P_{2})$  & $0.19$  &  $0.15$  &  $0.15$ \\
  &  $D_{s1}(2P_{1})$  & $6.80$  &  $6.47$  &  $6.47$  \\
  &  $D_{s1}(2P_{1}^\prime)$  & $6.56$  &  $6.40$  &  $6.40$ \\
  &  $D_{s2}^{\ast}(2^{3}P_{2})$  & $0.12$  &  $0.12$  &  $0.12$ \\
$D_{s2}(2D_{2})$  &  $D_{s1}(1P_{1}^\prime)$  & $5.37$  &  $5.61$  &  $5.61$  \\
  &  $D_{s1}(1P_{1})$  & $0.31$  &  $0.22$  &  $0.22$ \\
  &  $D_{s2}^{\ast}(1^{3}P_{2})$  & $0.48$  &  $0.28$  &  $0.28$ \\  &  $D_{s1}(2P_{1}^\prime)$  & $11.93$  &  $11.37$  &  $11.37$  \\
  &  $D_{s1}(2P_{1})$  & $1.98$  &  $1.94$  &  $1.94$ \\
  &  $D_{s2}^{\ast}(2^{3}P_{2})$  & $1.31$  &  $1.29$  &  $1.29$\\
$D_{s3}(2^{3}D_{3})$ &  $D_{s2}^{\ast}(1^{3}P_{2})$ &  $0.002$  &  $0.07$  & $0.09$  \\
  &  $D_{s2}^{\ast}(2^{3}P_{2})$ &  $8.52$  &  $8.34$  & $8.34$  \\
\bottomrule[1pt]
\end{tabular}
\end{table}

\section{Summary}
\label{Sec:Sum}
The radiative decay is one of important decay modes of charmed strange mesons, especially for the low lying charmed strange mesons. In the present work, we adopt a relativistic constituent quark model to depict the mass spectroscopy of the charmed meson, in which $D_{s0}(2317)$ and $D_{s1}(2460)$ are considered as $1^3P_0$ and $1P_1^\prime$ charmed strange mesons, respectively, while $D_{sJ}(3040)$ is assigned as $D_{s1}(2P_1)$ states.

With the wave function estimated by the relativistic quark model, we evaluate the electric transitions between the charmed strange mesons. By comparing the transition widths obtained with different approximations, we find that the long wave length approximation is reasonable for most cases of the electric transitions between charmed-strange mesons. Our estimation indicates that the partial widths of $D_{s0}(1^3P_0) \to D_s^\ast \gamma$, $D_{s1}(1P_1) \to D_s^\ast \gamma$ and $D_{s1}(1P_1) \to D_s \gamma$ are all safely under the upper limits of the experimental data. As for $D_{s1}(1P_1^\prime) \to D_s \gamma $ and  $D_{s1}(1P_1^\prime) \to D_s^\ast \gamma $, our estimation find that the branching ratios of these processes are large enough to be detected, which could be searched in further experiments in Belle II and LHCb. As for $P\to D\gamma$ and $D\to P \gamma $ processes, the width of some channels can reach up to 10 keV, which may be tested by further experimental measurements.

\section*{Acknowledgement}
This work is supported by the National Natural Science Foundation of China (NSFC) under Grant No. 11775050 and 11975075.

\appendix
\section{Electromagnetic transition operator}
\label{Sec:App}

By replacing the quark field $\bar{\psi}$ with $\psi^\dagger$, one can use matrix $\alpha$ instead of the $\gamma$ matrix in Eq. (\ref{Eq:He}). Then the electromagnetic transition matrix elements for a radiative decay process becomes, 
\begin{eqnarray}
	\mathcal{M}&=& \left \langle f \left| \sum_{j} e_j \alpha_j \cdot \epsilon e^{-i \mathbf{k}\cdot \mathbf{r}_j} \right| i \right\rangle\nonumber\\
	\end{eqnarray}
Considering the fact that the involve mesons are composite systems and the relativistic Hamiltonian is,
\begin{eqnarray}
	\hat{H}=\sum_j \left(\mathbf{\alpha}_j \cdot \mathbf{p}_j +\beta_j m_j \right) +\sum_{i,j} V\left({\mathbf{r}_i-\mathbf{r}_j}\right),
\end{eqnarray}
we have the following identity,
\begin{eqnarray}
	\mathbf{\alpha}_j\equiv i \left[ \hat{H},\mathbf{r}_j\right].
\end{eqnarray}
Then, the electromagnetic transition matrix can be expressed as,
\begin{eqnarray}
	\mathcal{M}&=& i \left\langle f\left| \left[\hat{H},\sum_j e_j \mathbf{r}_j\cdot \mathbf{\epsilon} e^{-i\mathbf{k} \cdot \mathbf{r}_j} \right]\right|i \right\rangle\nonumber\\
	&+& i \left\langle f\left| \sum_j e_j \mathbf{r}_j\cdot \mathbf{\epsilon} \mathbf{\alpha}_j \cdot \mathbf{k} e^{-i\mathbf{k} \cdot \mathbf{r}_j} \right|i \right\rangle\nonumber \\
	&=& -i(E_i-E_f-\omega_\gamma) \left \langle  f\left|g_e\right|i\right \rangle -i\omega_\gamma \left \langle  f\left|h_e\right|i\right \rangle, \label{Eq:App-M4}
\end{eqnarray}
with
\begin{eqnarray}
	h_e&=&\sum_j e_j \mathbf{r_j} \cdot \mathbf{\epsilon} (1-\mathbf{\alpha}_j \cdot \hat{\mathbf{k}}) e^{-i\mathbf{k} \cdot \mathbf{r}_j},\nonumber\\ g_e&=&\sum_j e_j \mathbf{r_j} \cdot \mathbf{\epsilon} e^{-i\mathbf{k} \cdot \mathbf{r}_j}. 
\end{eqnarray}
$E_i$, $E_f$ and $\omega_\gamma$ in Eq.~ (\ref{Eq:App-M4}) are the energies of the initial meson, the final meson and the emitted photon, respectively. Thus, $E_i-E_f-\omega_\gamma \equiv 0$ due to the conservation of energy.  Thus, one has, 
\begin{eqnarray}
	\mathcal{M}=-i\omega_\gamma \left \langle  f\left|h_e\right|i\right \rangle
\end{eqnarray} 
Following the procedures used in Refs.~\cite{Brodsky:1968ea, Li:1997gd}, one can get the non-relativistic expansion of $h_e$, which is,
\begin{eqnarray}
	h_e \simeq \sum_j \left[e_j \mathbf{r}_j \cdot \mathbf{\epsilon} -\frac{e_j}{2m_j} \mathbf{\sigma}_j \cdot \left( \mathbf{\epsilon \times \hat{\mathbf{k}}}\right)\right] e^{-i\mathbf{k} \cdot \mathbf{r}_j},
\end{eqnarray}
where the first and the second terms are corresponding to electric and magnetic transitions, respectively.

\end{document}